\documentclass{article}

\PassOptionsToPackage{numbers, compress}{natbib}

\usepackage[preprint]{neurips_2023}

\usepackage[utf8]{inputenc} 
\usepackage[T1]{fontenc}    
\usepackage{hyperref}       
\usepackage{url}            
\usepackage{booktabs}       
\usepackage{amsfonts}       
\usepackage{nicefrac}       
\usepackage{microtype}      
\usepackage{xcolor}         
\usepackage{graphicx}
\usepackage{textcomp}
\usepackage{subcaption}

{}
{}

\title{AI4GCC - Team: Below Sea Level \\ Score and Real World Relevance}

\author{%
Phillip Wozny*
\\
Tilburg University \\ Vrije Universiteit Amsterdam 
\And
Bram Renting*
\\
Leiden University \\ Delft University of Technology
\AND 
Robert Loftin \\
Delft University of Technology
\And 
Claudia Wieners\\
Utrecht University
\And
Erman Acar \\
University of Amsterdam 
}

\begin{document}

\maketitle

\begin{abstract}
As our submission for track three of the AI for Global Climate Cooperation (AI4GCC) competition, we propose a negotiation protocol for use in the RICE-N climate-economic simulation. Our proposal seeks to address the challenges of carbon leakage through methods inspired by the Carbon Border Adjustment Mechanism (CBAM) and Climate Clubs (CC). We demonstrate the effectiveness of our approach by comparing simulated outcomes to representative concentration pathways (RCP) and shared socioeconomic pathways (SSP). Our protocol results in a temperature rise comparable to RCP 3.4/4.5 and SSP 2. Furthermore, we provide an analysis of our protocol's World Trade Organization compliance, administrative and political feasibility, and ethical concerns. We recognize that our proposal risks hurting the least developing countries, and we suggest specific corrective measures to avoid exacerbating existing inequalities, such as technology sharing and wealth redistribution. Future research should improve the RICE-N tariff mechanism and implement actions allowing for the aforementioned corrective measures. 
\end{abstract}

\section{Introduction}

The 2022 UN Emissions Gap Report \cite{unep} indicates that states are not on track to keep global average surface temperature rise under the Paris Agreement target of 1.5\textdegree C since preindustrial levels \cite{pohl2023enjoying}. The frequency of extreme weather events increases exponentially with temperature rise. For instance, the frequency of unseasonably hot days increases six times under 1\textdegree C temperature rise and 20 times under 2\textdegree C temperature rise. As such, failing to meet the 1.5\textdegree C target would have disastrous consequences \cite{knutti2016scientific, fischer2015anthropogenic}.

The difficulties of addressing climate change and potential avenues for building international cooperation become apparent after reframing the problem as a public good social dilemma~\cite{rashidi2022strategic}. The climate is a global public good as all states benefit from its well-being even if they are not directly responsible for its maintenance through emission reduction policies. As such, states face the dilemma of whether to rationally pursue their self-interest or to altruistically cooperate to maintain the climate~\cite{kaul1999defining,kaul1999global}. In this context, free-riding implies benefiting from the emissions reductions of others while domestically failing to follow suit. Free-riding can also occur temporally, as the cost of climate damage is borne by subsequent generations. Policymakers are tasked with the design of mechanisms that alter rewards such that cooperation becomes the rational choice~\cite{nordhaus2015climate}.

Multi-agent systems (MAS) is a framework suited for modeling social dilemmas~\cite{de2006learning, leibo2017multi}. MAS aids policymakers by shedding light on the decision-making processes of simulated agents~\cite{zhang2022ai}.

In the present policy brief, we will leverage MAS as a means of discovering climate negotiation protocols. We will identify which features are optimal with respect to both climate and economic health and devise a list of policy recommendations based on the optimal protocol. Finally, we will characterize the feasibility and ethical considerations of implementing the suggested policy.

\section{The shortcomings of previous policies}
The 1997 Kyoto Protocol aimed to reduce greenhouse gas (GHG) emissions in developed countries to 5\% less than pre-1990 levels~\cite{gardiner2004global}. Relying on voluntary action and without sanction mechanisms, the Kyoto Protocol failed to meet its goal~\cite{nordhaus2020climate}. Except for eight of the 15 EU signatories, developed countries failed to meet their targets. Relative to pre-1990 levels, Canada increased GHG emissions by 25\%, Japan by 14\%, and the United States by 8.4\%~\cite{jakob2021carbon}.

Subsequent attempts to reduce GHG emissions enabled free-riding due to the following policy conditions. In 2004, the European Union (EU) established its Emissions Trading System (ETS), in which businesses can purchase emission offset certificates, thereby increasing the cost of carbon in participating countries~\cite{schippers2022proposal}. The 2015 Paris Climate agreement allowed states to set their own cost of carbon, resulting in carbon leakage~\cite{overland2022climate, branger2016carbon, monjon2011border}. The aforementioned occurs when emissions increase in GHG emissions in one country as a result of emission decreases in another country. Carbon leakages have two principal causes, demand for fossil fuels by non-mitigating countries and competitive relocation of carbon-intensive industries to save costs. Carbon-leaking countries are free-riders as they do not pay the cost of emissions reduction; thereby, lowering their prices and increasing competitiveness~\cite{jakob2021carbon}. 

Multiple international agreements aim to address carbon leakage. The Carbon Border Adjustment Mechanism (CBAM), a feature of the European Green Deal taking effect in 2026, imposes a tariff on carbon-intensive goods whose value depends on the difference between the EU and the exporting countries' carbon taxes. That is, the import price of goods from carbon-leaking states will increase to levels comparable to that of ETS complaint states. Carbon-leaking countries can only lower their tariff by taxing carbon domestically, thereby reducing their emissions~\cite{pirlot2022carbon, vidigal2022false, overland2022climate}. However, by narrowly focusing on exports, CBAM leaves large swaths of carbon-intensive economic activity untouched~\cite{tarr2022carbon}. Initially proposed by William Nordhaus and codified by Article 6 of the Paris Agreement, a carbon club (CC) is a similarly constructed mechanism centered around a coalition of states that agree to a common emissions reduction target and impose a uniform tariff on all goods whose value depends on the domestic carbon tax of an exporting country. Therefore countries that export non-carbon-intensive goods but still fail to reduce GHG domestically will face higher tariffs than club members. Said exporter can always join the club by taxing carbon domestically and reducing emissions~\cite{nordhaus2015climate, nordhaus2021dynamic}.

\section{Proposed solution and recommendations}

Track two of the AI4GCC competition invites participants to develop and evaluate novel mechanisms within the RICE-N climate and economic model~\cite{zhang2022ai}. Said mechanisms take the form of negotiation protocols, and sets of actions independent of the climate-economic simulation step. Agreements between states made during the negotiation steps constrain the space of actions possible during the climate-economic simulation step. 

We developed a base negotiation protocol that borrows the CBAM and CC tariff mechanism called \emph{Basic Club} (BC) which works as follows. 
\begin{itemize}
    \item Once every five years, states gather to propose a target mitigation level.
    \item States accept or reject the proposed mitigation levels and commit to the highest accepted mitigation level, thereby always mitigating at least the given level. A club is formed as the subset of states committed to a given mitigation level.
     \item Club members set a minimum tariff on goods from states with mitigation rates lower than the club's. The tariff minimum inversely depends on the mitigation rate of the exporting country. For example, a club of mitigation 9 will, to a member of club 7, tariff at least 3. 
    \item Club members set a maximum tariff limit for states with mitigation rates greater than or equal to theirs. The tariff ceiling depends inversely on the mitigation rate of the exporting country. For example, a club of mitigation 6 will, to a member of club 8, tariff at most 2. 
    \item The climate and economy and simulated for a five-year period and the process repeats. 
\end{itemize}

\begin{table}
\caption{The design elements employed to modify the base protocol. Discrete Defect (DD) resulted in the best balance of economic and climate performance.}\label{tb:de}
\resizebox{\textwidth}{!}{
\begin{tabular}{p{0.35\linewidth} p{0.65\linewidth}}
\toprule
\textbf{Name}                      & \textbf{Description}                                                                                                                           \\ \midrule
Discrete Defection (DD)            & Discrete Defection adds an action step called Defect. Defecting states are no longer obliged to mitigate at the level of their club. \\ \midrule
Free Trade (FT)                    & Club members and states with mitigation rates that are higher than a given club have no tariffs. This is intended to incentivize the club membership further.  \\ \midrule
Max Punishment (MP)                & If a state defects once, then it receives a maximum tariff for the rest of the simulation and loses access to free trade zones.         \\ \midrule
Hard Defect (HD)                   & Defecting results in zero-level mitigation for a single economic step.                                                                
\\ \bottomrule
\end{tabular}}
\end{table}

We modified the foundational BC protocol with specific \emph{design elements} (see \autoref{tb:de}) intended to capture specific phenomena, such as states which are nominally committed to emissions reduction but in practice fail to follow through. By evaluating modified negotiation protocols along economic or climate dimensions, measured as \emph{gross output} and \emph{temperature rise}, respectively, we can explore the Pareto frontier of optimal design elements. Furthermore, we can illuminate how specific design elements work towards either climate or economic goals.

\begin{figure}
    \centering
    \includegraphics[scale=.7]{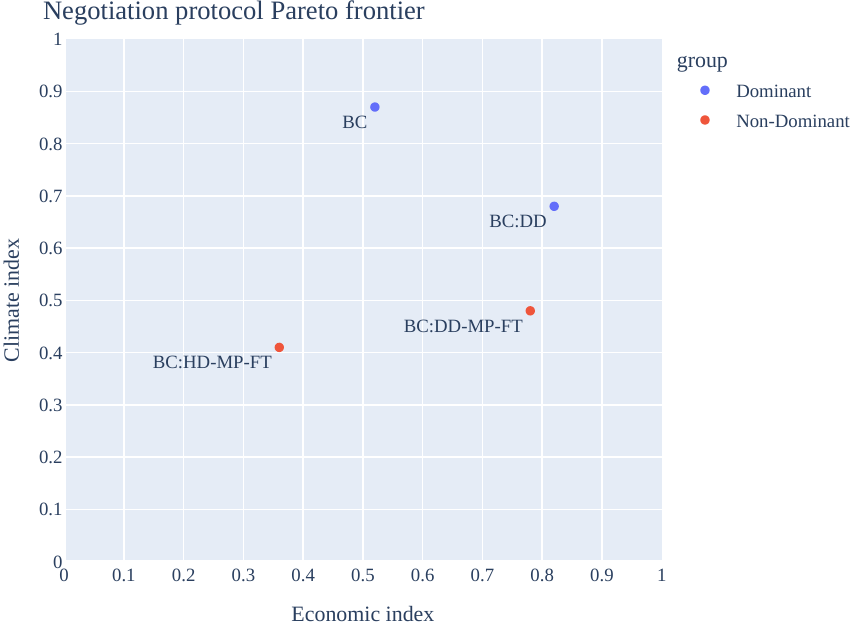}
    \caption{A comparison of the economic and climate metrics for multiple variations on the BC protocol. The coloration of points indicates whether the particular protocol is Pareto dominant or not.}
    \label{fig:pareto}
\end{figure}

Comparing different combinations of the design elements, \autoref{fig:pareto} suggests that the optimal ones are Discrete Defect (BCDD) and BC without any modifications. BCDD and BC differ in how they handle defection. With BCDD, states can propose ambitious mitigation targets, then decide not to follow through as a discrete action. Under BC, on the other hand, states can only defect by proposing or accepting low mitigation rates. BCDD is the negotiation protocol that best balances economic and climate metrics. However, BC results in better climate outcomes. We believe that the difference in economic index between BC and BCDD is not an accurate representation of the real-world damage that would result from approximately 1\textdegree C temperature rise (see submission 3 for our critique of the damage function). As such, we base our recommendations on BC. A formal description of the protocol can be found in Appendix B. 


\paragraph{Policy recommendations}

Our solution can be implemented through the following policy recommendations:

\begin{itemize}
    \item Allow states with common mitigation ambitions to establish CCs. 
    \item Let each CC impose a uniform tariff on all goods from non-club members whose value depends on the difference between the exporting country and the CC's emission costs~\cite{nordhaus2015climate}.
    \item Create free trade zones within CCs and between CCs of higher mitigation levels~\cite{hovi2016climate, sabel2017governing}.
    \item Redistribute tariff revenue to support developing countries negatively impacted by the CC and finance sustainable infrastructure~\cite{nordhaus2021dynamic, perdana2022making}.
\end{itemize}

\section{Effectiveness} 

After training our model according to the configuration described in Appendix A, we compared our proposed protocol to No Protocol, which consists of disabling all negotiation functionality in the RICE-N model. \autoref{table:rcp} compares the evaluated protocols to their corresponding representative concentration pathways (RCP) and shared socioeconomic pathways (SSP) based on estimated temperature rise. Both pathways represent possible emissions scenarios based on a given degree of global mitigation~\cite{meinshausen2020shared, masson2021climate}.

\subsection{Comparison to RCP and SSP pathways}

\begin{figure}
    \centering
    \includegraphics[width=\textwidth]{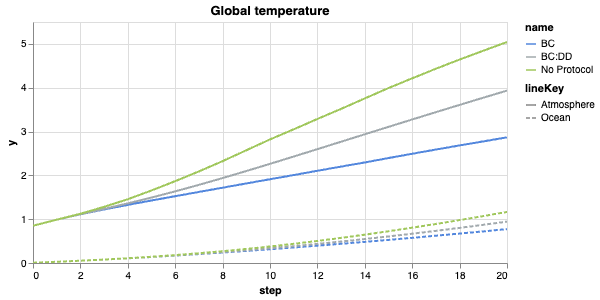}
    \caption{Global temperatures of our Pareto dominant protocols compared to No Protocol.}
    \label{fig:gt}
\end{figure}

By comparing temperature rise from a protocol to a pathway, we can make inferences about the resulting climate, demographic and economic consequences of a given protocol. As visible in \autoref{fig:gt}, our proposal results in a temperature rise of 2.09\textdegree C, compared to a 4.43\textdegree C rise of no protocol benchmark. As such, its resulting climate and economic conditions are comparable to RCP 3.4/4.5 and SSP 2.

\begin{table}
\centering
\caption{Our proposed protocol aligns with RCP 3.4 and SSP 2 in approximate temperature rise over a 100-year time period. Both scenarios are considered compromises between extremely stringent and overly lenient approaches.}\label{table:rcp}
\resizebox{\textwidth}{!}{
\begin{tabular}{llll}
\toprule
Protocol Name     & Temperature Rise & Corresponding RCP  & Corresponding SSP \\ \midrule
No Protocol       & 4.43             & RCP 7.5/8.5      & SSP 7         \\
BCDD & 3.20 &  RCP 6.0 & SSP 2/4.5 \\
BC & 2.09             & RCP 3.4/4.5      & SSP 2        \\
\bottomrule
\end{tabular}}
\end{table}

RCP and SSP pathways are evaluated by the degree to which they change the frequency of weather events that previously occurred once in 10 years in the period between 1850-1900. For example, with SSP 2, once in 10 years extreme heat events will likely occur 5.6 times under our protocol compared to 9.4 times under the benchmark. Heavy precipitation one in 10-year events will occur 1.7 times under our protocol compared to 2.7 times without a protocol. One in 10-year agricultural and ecological droughts will increase frequency from once every 10 years to 2.4 and 4.1 occurrences per 10 years under our protocol and the benchmark, respectively~\cite{masson2021climate}.

By 2100, sea levels would rise between .66 and 1.33 meters under our protocol compared to the benchmark's rise between .98 and 1.88 meters. Moreover, drought-induced migration is estimated to increase by an average of 201\% under the proposed protocol compared to 477.4\% globally\cite{smirnov2023climate}. 

As evident in the comparison to aforementioned pathways, our protocol is preferable to complete inaction; however, the resulting climate conditions would still result in drastic changes in the frequency of extreme weather events, sea level rise, and mass migration ~\cite{masson2021climate, smirnov2023climate, meinshausen2020shared}. This underscores the necessity of a swift and coordinated response to climate change. 


\begin{figure}
    \centering
    \begin{subfigure}[t]{0.49\linewidth}
        \centering
        \includegraphics[width=\linewidth]{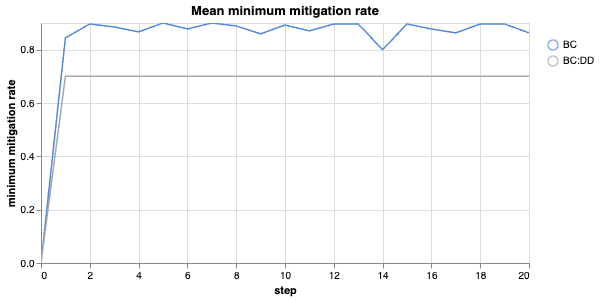}
        \caption{Average mitigation rate of all regions. Note, No Protocol does not feature here as the minimum mitigation rate is a negotation specific value.}\label{fig:mmr}
    \end{subfigure}
    \hfill
    \begin{subfigure}[t]{0.49\linewidth}
        \centering
        \includegraphics[width=\linewidth]{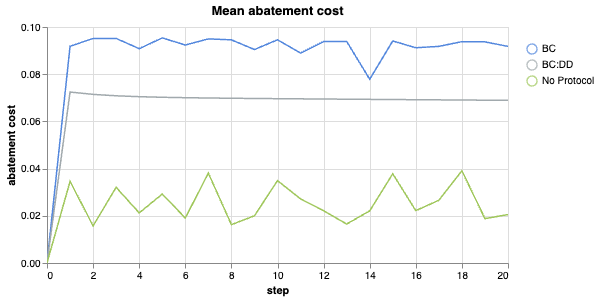}
        \caption{Average abatement cost of all regions}\label{fig:mac}
    \end{subfigure}
    \caption{}
\end{figure}

The average mitigation rates and associated abatement costs are illustrated in \autoref{fig:mmr} and \autoref{fig:mac}. Assuming that our protocol corresponds to SSP~2, then our protocol would likely result in a disproportionate financial burden on developing countries and fossil fuel-dependent regions~\cite{leimbach2019burden}. Though the RICE-N simulated regions do not directly map onto real-world correlates, we can evaluate the correlation of total abatement costs with features indicative of developing countries; namely, capital, production factor, carbon intensity, and gross output. For this purpose, we evaluated BC:DD 40 times to generate sufficient output data to measure the aforementioned correlations. 

\begin{figure}
    \centering
    \begin{subfigure}{0.49\textwidth}
        \centering
        \includegraphics[width=\textwidth]{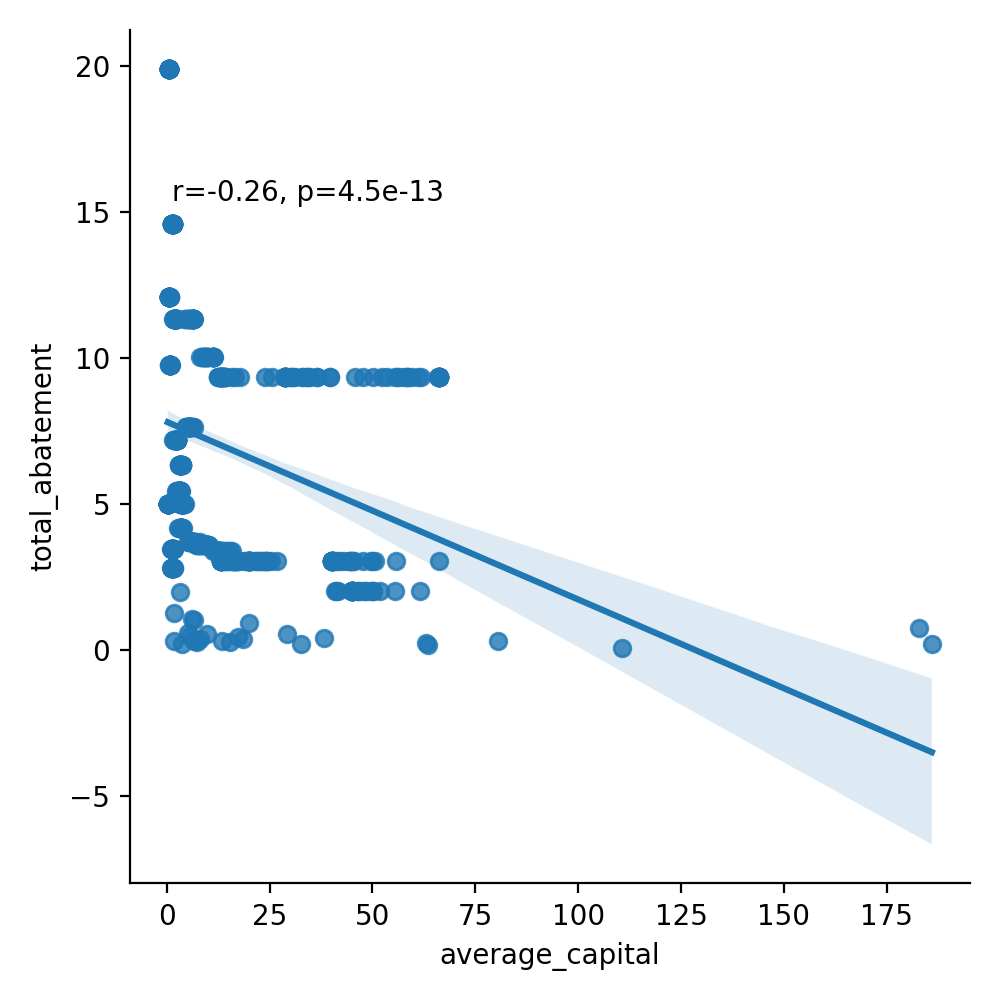}
    \end{subfigure}
    \hfill
    \begin{subfigure}{0.49\textwidth}
        \centering
        \includegraphics[width=\textwidth]{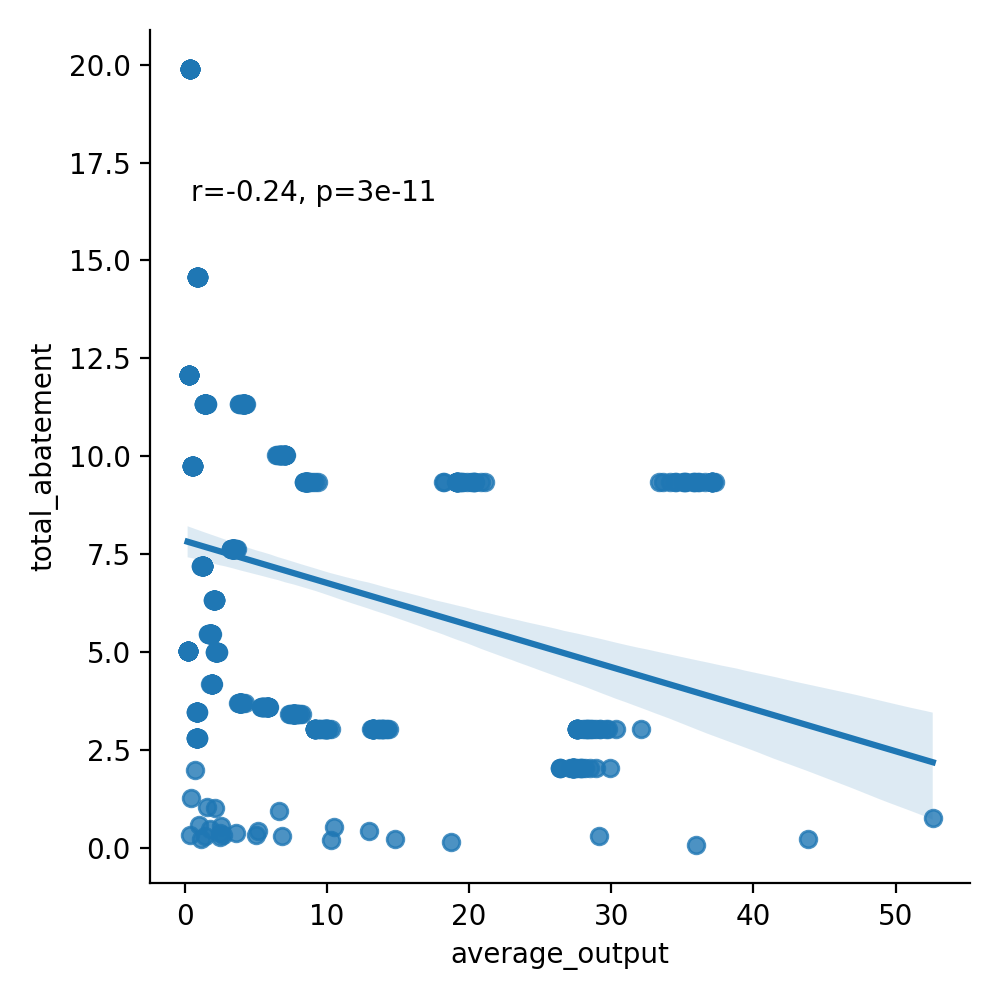}
    \end{subfigure}
    
    \begin{subfigure}{0.49\textwidth}
        \centering
        \includegraphics[width=\textwidth]{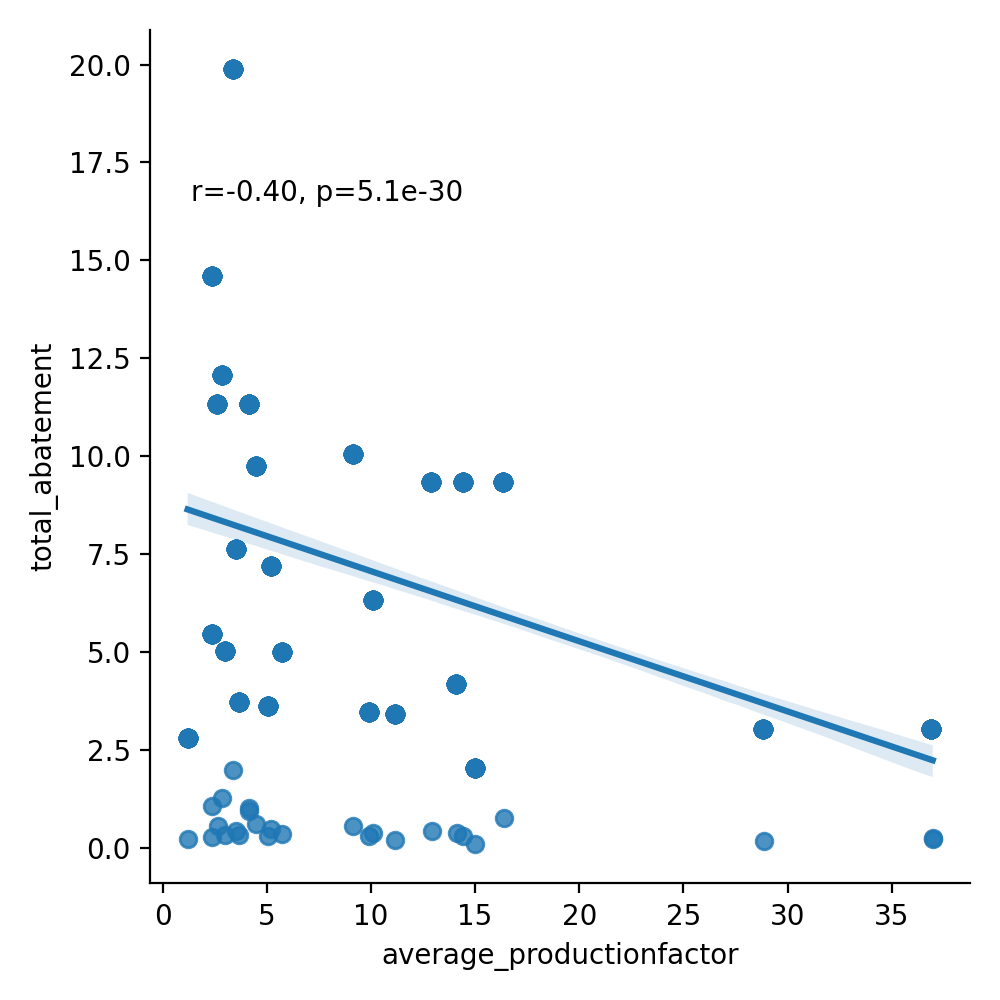}
    \end{subfigure}
    \hfill
    \begin{subfigure}{0.49\textwidth}
        \centering
        \includegraphics[width=\textwidth]{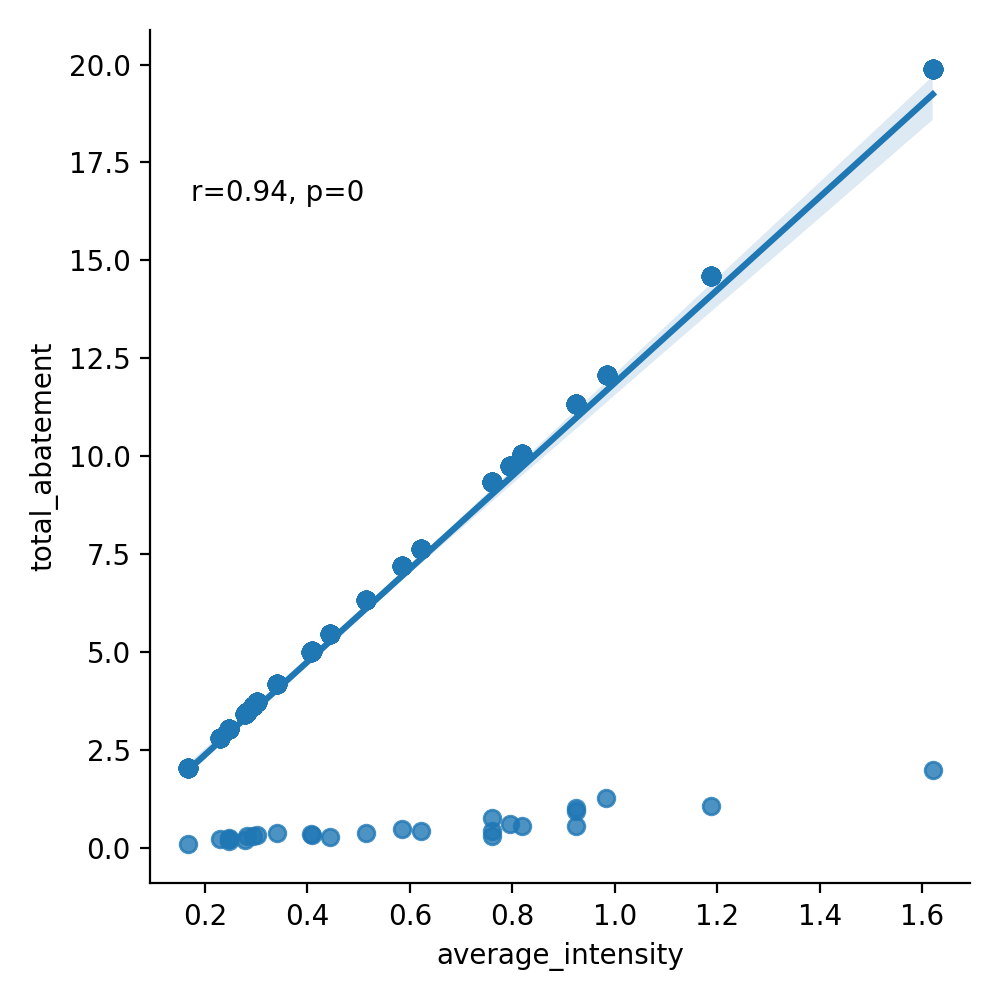}
    \end{subfigure}
    \caption{Pearson correlation coefficient between abatement and capital, output, carbon intensity, and production factor.}\label{fig:cor}
\end{figure}

As is shown in \autoref{fig:cor}, abatement is slightly negatively correlated with both capital and output ($r=-0.26$, $r=-0.25$, respectively). To a small but significant degree ($p<.05$), low-capital states pay more for climate mitigation. Production factor, a measure of the level of technology of a given state, is moderately negatively correlated with abatement costs ($r=-.4$, $p<.05$); that is, high-technology states tend to pay less for climate mitigation. Furthermore, carbon intensity is very strongly positively correlated with abatement costs ($r=0.94$, $p<.05$). This intuitive result indicates that states whose economic activity depends on carbon pay significantly more in mitigation costs. 

The imbalance of abatement burden sharing underscores the necessity of technology transfer and wealth redistribution policies as an accompaniment to our proposed protocol. However, in its current state, such measures are outside the scope of the current RICE-N implementation.

\section{Feasibility}

\paragraph{Legal concerns}

The chief concern regarding Carbon Border Adjustment Mechanism (CBAM) adjacent legal instruments is their compliance with the World Trade Organization (WTO) General Agreement on Tariff And Trade's (GATT) ``most favored nation'' clause, stating that tariffs must be non-discriminatory. That is, a tariff applied to a single good must be the same for all states exporting it. While CBAM would seem to violate that clause, exceptions are made in the following conditions~\cite{vidigal2022false}. 

\begin{itemize}
    \item The agreement promotes one of the GATT article XX (g) objectives; namely, ``relating to the conservation of exhaustible natural resources.''
    \item The agreement should contribute to the objective.
    \item The agreement should not discriminate between countries. If it appears to, then its discrimination must be on the grounds justifies the rationale. 
\end{itemize}

This legal framework has precedent since the 1998 WTO Appellate Body Report ``United States - Import Prohibition of Certain Shrimp and Shrimp Productions''~\cite{shaffer1999united}. Our proposed protocol could then inherit CBAM's WTO compliance with respect to non-discrimination. 

Uniform tariffs are the other key component of our solution. The case for WTO compliance with uniform tariffs is grounded in two ways. First, uniform tariffs can correct trade imbalances caused by carbon-leaking states who have an unfair advantage. Second uniform tariffs can be considered ``sanctions against misconduct'' as is standard practice in foreign relations~\cite{pihl2020climate, mavroidis201516}. 

\paragraph{Administrative concerns}

The advantage of a CC over CBAM is the administrative simplicity of a uniform tariff. The CBAM only applies to a subset of goods. Determining the carbon content of goods is costly, opaque, and vulnerable to fraud. Given the impact of tariffs on affected industries, carbon monitoring invites extensive corporate lobbying which further distorts the accuracy of an audit~\cite{bierbrauer2021co2}. In contrast, a uniform tariff does not require carbon content determination and does not seek to punish any specific industry. Rather, uniform tariffs promote CC participation by affecting all exports~\cite{nordhaus2015climate}. Therefore, CCs are administratively more feasible than CBAM and more robust against manipulation. However, both CCs and CBAM require states to calculate carbon pricing. Some developing states lack the administrative apparatus necessary to calculate and tax carbon emissions domestically \cite{dadush2021eu}. Tariff revenue generated will need to be reinvested in carbon pricing administration of developing countries. 

\paragraph{Political Obstacles}

Whether CBAM and similar policies are successful depends partially on the involvement of large states, such as the US. In the US, climate ambitions are highly politicized, as evident in former president Trump's withdrawal from the Paris Climate Agreement \cite{martin2019multi}. Moreover, the US does not have a national ETS \cite{dadush2021eu}. Despite wavering commitments at the federal level, sub-national coalitions of states remain committed to international climate agreements through regional carbon markets. Furthermore, states politically opposed to climate mitigation may still be convinced to join the CBAM due to trade dependencies with the EU \cite{martin2019multi}.

\section{Ethical considerations \& risks}

Structurally similar to CBAM, our proposal inherits its ethical considerations. Specifically, our  protocol risks reproducing existing developmental inequalities unless specific corrective measures are taken into account \cite{perdana2022making}. The least developing countries (LDC) that trade with the EU yet lack green industry will be disproportionately hurt by CBAM and thereby by our protocol as well. Given that offering exemptions to CBAM invites further carbon leakage, LDCs will require both revenue redistribution and technology sharing \cite{perdana2022making, nordhaus2021dynamic}. The former can weaken the impact of tariffs temporarily while the latter aids the LDC in increasing domestic mitigation efforts. Financing technology transfer and revenue redistribution are possible if tariff revenue is reallocated to these ends. Revenue raised by tariffs on LDC imports can be reinvested back in the LDCs themselves \cite{eicke2021pulling}. Failing to reinvest tariff revenue into technology transfer and revenue redistribution risks turning CBAM into a simple tax on LDCs \cite{goldthau2022open}. 

There is also the ethical risk of a literal interpretation of the model outputs. There is a simulation to reality gap that must be acknowledged by policymakers who intend to leverage its insights. Furthermore, there are concerns that CBAM and related policies can introduce further trade distortions \cite{lim2021pitfalls}. However, our proposal intends to correct trade distortions caused by carbon leakage ~\cite{bierbrauer2021co2}.

\section{Conclusion}

Carbon leakage undermines previous policy attempts to curtail GHG emissions globally \cite{overland2022climate}. Drawing inspiration from the extensive literature addressing carbon leakage, we leverage reinforcement learning in a multi-agent system to simulate novel negotiation protocols which cultivate international cooperation. Our proposed policy, results in climate and economic conditions comparable to RCP 3.4/4.5 and SSP 2. While these are by no means the worst socioeconomic pathways possible, they still imply dramatic increases in the frequency of extreme weather events and come at great economic cost \cite{masson2021climate, smirnov2023climate}. Furthermore, our proposal runs the risk of punitively taxing developing countries lacking in green infrastructure. As such, it is critical that technology is shared and wealth is redistributed to avoid exacerbating existing inequalities \cite{goldthau2022open}. 

One shortcoming of our present submission is the lack of robustness testing. This can be performed by measuring the resilience of clubs to defection by principal members, states with the highest output, and their trading partners. However, as evident in submission 3, the sanction mechanism of RICE-N does not impact reward as expected. Correcting this is a necessary precondition to proper robustness testing. See submission 3 for details. 

Future research on RICE-N for modeling cooperation should correct the tariff component, allow for technology sharing and wealth redistribution, and perform thorough robustness testing. Furthermore, RICE-N can be directly calibrated to real world nation-states for more realistic scenario analysis, such as resilience of climate clubs in the face of defection by large states.

\begin{ack}
We would like to thank Maikel van der Knaap, Cale Davis, Albert Bomer, Catholijn Jonker, and Holger Hoos for their time spent discussing various topics of this competition.

This research was (partly) funded by the \href{https://hybrid-intelligence-centre.nl}{Hybrid Intelligence Center}, a 10-year programme funded by the Dutch Ministry of Education, Culture and Science through the Netherlands Organisation for Scientific Research, grant number 024.004.022.
\end{ack}

\bibliographystyle{IEEEtranN}

\bibliography{ref}

\end{document}


\begin{appendices}

\section{Training}
\label{appendix:training}

\begin{table}[H]
\centering
\caption{Training configuration}\label{tab:config}
\begin{tabular}{ll}
\toprule
Parameter                      & Value                                \\ \midrule
number episodes                & 100000                               \\
batch size in episodes         & 60                                   \\
framework                      & torch / rllib                        \\
value function loss coefficient & 0.1                                  \\
entropy coefficient schedule   & (0, 0.5), (40000, .1), (70000, 0.05) \\
clip grad norm                 & TRUE                                 \\
max grad norm                  & 0.5                                  \\
gamma                          & .92                                  \\
learning rate                  & 0.0005                               \\ \bottomrule
\end{tabular}
\end{table}

\section{Formalization}
\label{appendix:formalization}
For the negotiation protocols referenced in this submission, we utilized the configuration referenced in \autoref{tab:config}. Furthermore, we utilized a modified masking mechanism in the model to mask out all actions not relevant to a given step. We do this to stabilize training.

In addition to the formalism stated in the original white paper, given $n$ agents (i.e., $[n] = \{1, \ldots, n\}$)\footnote{In the simulations, $n = 27$}, action space $k$ has two subspace crucial to our  protocol 
\begin{itemize}
    \item  \textbf{proposal} $p_i \in \{0, \ldots, 9\}$ which corresponds to the  mitigaton level  that every agent $i \in [n]$ chooses for the climate club it takes place. 

    \item \textbf{evaluation} $\mathbf{e}_i =   \langle e_1, \ldots, e_{n} \rangle$   is a vector representing \emph{evaluation} of agent $i$ for all the other agents; that is,  each entry $e_j \in \{0, 1\}$ (with $j \in [n]$) denotes that the agent $i$ either accepts (i.e., $e_j = 1$) or rejects (i.e., $e_j = 0$) the proposal from agent $j$. \footnote{Here, to keep things simple, we ignored the case where $i=j$.}
\end{itemize}

\end{appendices}